# Low-threshold efficient $N_2^+$ lasing driven by sub-cycle soliton dynamics in a hollow waveguide

Tiandao Chen[1], Zhiyuan Huang[1,2,3,*], Jinyu Pan[1], Donghan Liu[1,2,3], Pengtao Wang[2,3], Xinglin Zeng[2,3], Jinxin Zhan[2,3], Jiapeng Huang[2,3], Wenbin He[2,3], Xin Jiang[2,3], Huailiang Xu[4], Yi Liu[5], Meng Pang[1,2,3,6,*], Yuxin Leng[1,6,*] and Ruxin Li[1,7]

[1]State Key Laboratory of Ultra-intense Laser Science and Technology, Shanghai Institute of Optics and Fine Mechanics (SIOM), Chinese Academy of Sciences (CAS), Shanghai 201800, China

[2]Russell Centre for Advanced Lightwave Science, Shanghai Institute of Optics and Fine Mechanics (SIOM) and Hangzhou Institute of Optics and Fine Mechanics (HIOM), Hangzhou 311400, China

[3]Zhejiang Key Laboratory of Microstructured Specialty Optical Fiber, Hangzhou Institute of Optics and Fine Mechanics (HIOM), Hangzhou 311400, China

[4]School of Optoelectronic Engineering, Xidian University, Xi'an 710071, China

[5]School of Optical-Electrical and Computer Engineering, University of Shanghai for Science and Technology, Shanghai 200093, China

[6]Hangzhou Institute for Advanced Study, University of Chinese Academy of Sciences, Hangzhou 310024, China

[7]Zhangjiang Laboratory, Shanghai 201210, China

[*]Corresponding author: huangzhiyuan@siom.ac.cn; pangmeng@siom.ac.cn; lengyuxin@siom.ac.cn

**The phenomenon of $N_2^+$ lasing, observed in femtosecond-laser filamentation, attract considerable interests in recent several years, with great application potentials in fields of remote sensing and ultrafast spectroscopy. Efficient $N_2^+$ lasing at relatively-low pump energies and with high beam quality, while being highly-demanded for applications, remains, however, quite challenging in practical experiments. Here, we demonstrate a new route of generating low-threshold $N_2^+$ lasing with unprecedently-high efficiency, which is enabled by soliton dynamics in a gas-filled hollow-tapered-capillary system. High-order-soliton compression of a 12-fs, 10-µJ-level pump pulse forms a sub-cycle asymmetric transient that tunnel-ionizes $N_2$ to $N_2^+$ and, through direct, single-photon resonant excitation, creates population inversion between the ground state $X^2\Sigma_g^+$ and the excited state $B^2\Sigma_u^+$—a dynamic process distinct from the widely adopted three-state coupling picture—and remarkably at unexpectedly low pump energy. In the experiments, we obtained 100-nJ-level $N_2^+$ lasing pulses at 391 nm with conversion efficiencies up to $3.3 \times 10^{-3}$, at pump energies of less than 50 µJ. These results represent improvement of more than one orders of magnitude in both generation efficiency and lasing threshold, compared with prevailing filamentation-based schemes. Our study bridges two generally-disparate fields (sub-cycle soliton dynamics and $N_2^+$ lasing), and paves the way for narrow-band, high-beam-quality lasing pulses that may find wide applications in advanced spectroscopy and nonlinear pump-probe experiments.**

The interaction of intense femtosecond laser pulses with gaseous media underpins both the exploration of new physics[1, 2] and the creation of advanced light sources[3-8]. One major direction exploits parametric nonlinear processes—primarily the third-order Kerr nonlinearity—to achieve spectral broadening, pulse compression, and frequency conversion[3, 9]. The other direction harnesses the quantized energy-level structure of gas atoms and molecules: strong-field tunnelling ionization, field-induced ionic state coupling, and population inversion can generate stimulated radiation amplification directly from the gas, offering a spectroscopic window into ultrafast molecular dynamics and enabling cavity-free gas lasers[10, 11]. The nitrogen-ion ($N_2^+$) lasing is a prime example of this second paradigm. According to the widely accepted three-state coupling picture[12, 13], when nitrogen molecules ($N_2$) are exposed to an intense, multi-cycle (> 10) femtosecond laser field, tunnel ionization from inner-valence orbitals populates $X^2\Sigma_g^+$, $A^2\Pi_u$, and $B^2\Sigma_u^+$ state of $N_2^+$, while post-ionization couplings between $X^2\Sigma_g^+$ and $A^2\Pi_u$ that enabled by the oscillating electric field in pulse trailing edge can evacuate the ionic ground state ($X^2\Sigma_g^+$) and establish population inversion between $B^2\Sigma_u^+$ and $X^2\Sigma_g^+$, giving rise to narrowband, coherent emission at visible (VIS) and ultraviolet (UV) wavelengths[12, 13]. Due to its remote generation capability, narrow linewidth, and cavity-free operation, $N_2^+$ lasing has attracted broad interest for applications from remote sensing to high-resolution pump-probe spectroscopy[14-17]. Nevertheless, the common free-space filamentation approach for $N_2^+$ lasing is fundamentally constrained: in general, forming a filament with multicycle pulses demands pulse energy of few millijoules, which restrict the conversion efficiency to $10^{-5}$-$10^{-4}$, and the high-spatial nonlinearity accumulation of the pump pulse can deteriorate the beam profile of the lasing signal[17]. A recent cascaded focusing scheme has raised the efficiency close to $10^{-3}$, yet without reducing the pump threshold below the millijoule level[17].

Gas-filled hollow-core fibres have recently emerged as a transformative platform that promises to overcome these limitations. By confining light and gas within a small circular core over an extended length, such waveguides can ensure a single mode propagation while dramatically enhance light–gas interaction, creating an ideal environment for extreme nonlinear processes and enabling high-quality light sources[1, 3-6, 18-26]. In particular, soliton dynamics in gas-filled hollow capillary fibres (HCFs) has drawn considerable attention: when a pump pulse with appropriate parameters propagates in the anomalous dispersion regime, the interplay of self-phase modulation (SPM) and negative group-velocity dispersion drives high-order soliton self-compression to sub-cycle domain, and simultaneously triggers the emission of broadly tunable UV dispersive waves (DW)[27]. Near the point of maximum temporal compression, the self-compressed pulse can reach a peak intensity sufficient to tunnel-ionize the filling gas[24, 28-31], creating a high-density $N_2^+$ column. More importantly, unlike conventional tens-of-femtosecond pump pulse that drives population dynamics of $N_2^+$ over multiple cycles, the self compressed pulses can have customized, few- or even sub-cycle waveforms[32], offering a window into new transient processes in gas lasing.

In this *Letter*, we report the first demonstration of sub-cycle soliton-dynamics-driven $N_2^+$ lasing. Using a $N_2$-filled tapered HCF, soliton self-compression of a 12-fs, 10-μJ-level near-infrared (NIR) pump pulse concurrently broadens its spectrum to 200 nm and ionizes nitrogen molecules to create a $N_2^+$ plasma column. Numerical simulations show that the self-compressed pulse can reach a duration of ~2 fs (less than one optical cycle of the driving laser), and its trailing edge carries a waveform with wavelength close to 391 nm, which resonant with the $B^2\Sigma_u^+ \rightarrow X^2\Sigma_g^+$ transition of $N_2^+$. As a result, this asymmetric sub-cycle transient rapidly redistributes the populations of the $X^2\Sigma_g^+$ and $B^2\Sigma_u^+$ states of $N_2^+$ produced by ionization, establishing pronounced population inversion within ~2 fs. Experimentally, we achieve fundamental-mode $N_2^+$ lasing signals with energies up to 163.2 nJ, conversion efficiencies up to $3.3 \times 10^{-3}$ and a pump energy threshold as low as 17 μJ, representing an improvement of more than one orders of magnitude in both metrics compared with conventional filamentation-based schemes. Our work establishes a new paradigm for efficient gas-laser generation by combining sub-cycle soliton dynamics with strong-field-induced $N_2^+$ lasing, a concept that could be readily extended to other matters for investigating light-matter interaction in sub-cycle domain or producing coherent radiation. Moreover, the resulting high efficiency, low threshold, high-beam-quality and narrow-band 391 nm pulses, which naturally collinear with UV DW pulses emitted by the driving soliton, are uniquely suited for single-beam coherent Raman scattering spectroscopy[14-16].

The soliton-driven $N_2^+$-lasing process is conceptually illustrated in Figs. 1a and 1b. When a high-order soliton propagates in the anomalous dispersion regime of an $N_2$-filled tapered HCF, it undergoes sub-cycle self-compression, reaching 100-TW/cm$^2$-level peak intensity. This intense laser field tunnel-ionizes nitrogen molecules, creating an $N_2^+$ column with density ~$10^{16}$ cm$^{-3}$, and establish population inversion between $B^2\Sigma_u^+$ and $X^2\Sigma_g^+$ within ~2 fs, thereby providing optical gain for $N_2^+$ lasing. In the frequency domain, the sub-cycle pulse corresponds to a 200-1200 nm supercontinuum[1, 33], which covers the $B^2\Sigma_u^+ \rightarrow X^2\Sigma_g^+$ transition and serves as the seed of the $N_2^+$-lasing process[12, 34, 35].

The experimental realization of the scheme employs a 450-mbar $N_2$ filled tapered HCF (core diameter decreases linearly from 160 to 96 μm in 65 cm length). In this hollow waveguide, the zero-dispersion wavelength decreases from 471 to 381 nm, exhibiting negative dispersion at wavelengths longer than 471 nm and therefore enables soliton self compression. The pump pulse is at ~800 nm with a ~12-fs pulse width and a 35-μJ pulse energy (soliton order $N \sim 4.3$), obtained from a commercial Ti:sapphire laser and a subsequent HCF compression system, see Methods and Supplementary note 1 for more details of the experimental set-up. The output spectrum measured at the HCF output port is illustrated in Fig. 1c, where sharp spectral lines at 358 nm and 391 nm on the supercontinuum are observed, confirming the $N_2^+$-lasing process. Thanks to the cylindrically symmetric geometry of the waveguide and the relatively-weak spatial nonlinearity accumulation of the pump pulse, these $N_2^+$-lasing signals exhibit perfect single-fundamental-mode purity and spatial coherence, see Fig. 1d and Supplementary note 1.

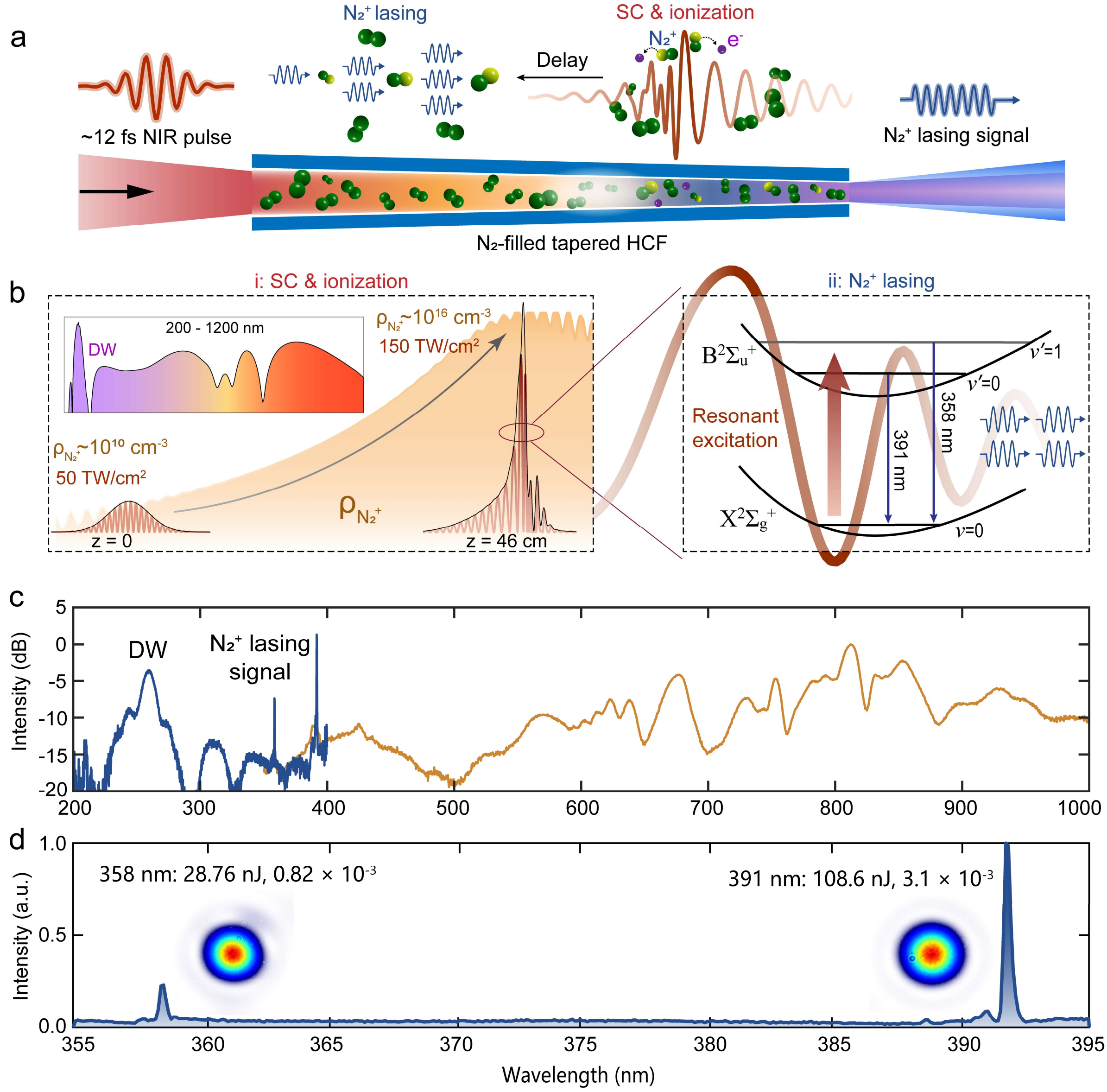

**Fig. 1 | Conceptual illustration of the soliton-driven $N_2^+$ lasing process and experimental results. a,b** Schematic of soliton-driven $N_2^+$ lasing in a $N_2$-filled tapered HCF. The top panel in (**a**) illustrates the dynamics of soliton sub-cycle compression (SC), strong-field ionization and $N_2^+$ lasing, while (**b**) illustrates these processes in detail. **c** Measured output spectrum from a tapered HCF with core diameter decreasing linearly from 160 to 96 μm and a length of 65 cm, filled with 450-mbar $N_2$, pumped by ~12-fs, 800-nm, 35-μJ pulse. The yellow and the blue lines are measured using a VIS-NIR spectrometer and a high-resolution UV spectrometer, respectively. DW, dispersive wave. **d** $N_2^+$ lasing spectrum and beam profiles, measured directly at the HCF output port.

In the experiment, we observed that the energy of the 391-nm $N_2^+$-lasing signal can reach beyond 100 nJ with only 35 μJ of pump energy, corresponding to ~3 × $10^{-3}$ conversion efficiency—an improvement of 1-2 orders of magnitude in both efficiency and pump threshold over conventional schemes[17, 35-37]. Surprisingly, the peak intensity at the beam centroid of the self-compressed pulse is estimated to be 150 TW/cm$^2$, a value lower than that in previous filamentation-based $N_2^+$-lasing experiment (usually higher than 200 TW/cm$^2$)[12, 13]. To uncover the underlying physics of this improvement, we simulated the pulse propagation in the tapered HCF using the unidirectional pulse propagation equation[3, 32, 38-40] to obtain the electric field at the maximum self-compression point, and studied the states-coupling process of $N_2^+$ in the self-compressed pulse by solving the time-dependent Schrödinger equation[12, 13] (see Supplementary note 2 for more details). The numerical results, shown in Figs. 2a and b, indicate that the pulse reaches maximum self-compression at 46.15 cm, with a peak intensity at the beam centroid of 150 TW/cm$^2$ and a pulse width of 2.1 fs. We assumed the tunneling ionization that initially populates $X^2\Sigma_g^+$, $A^2\Pi_u$ and $B^2\Sigma_u^+$ occurs at the peak of the electric field square, and the strong field-induced dynamics starts immediately after the ionization[12, 13]. This assumption has been widely adopted in simulations of few- or multi-cycle pulse-driven population dynamics in $N_2^+$ lasing[12, 13, 41-43], and is even better justified in our case, given the sub-cycle pulse can confine the ionization event to a narrow temporal window at the peak of the field.

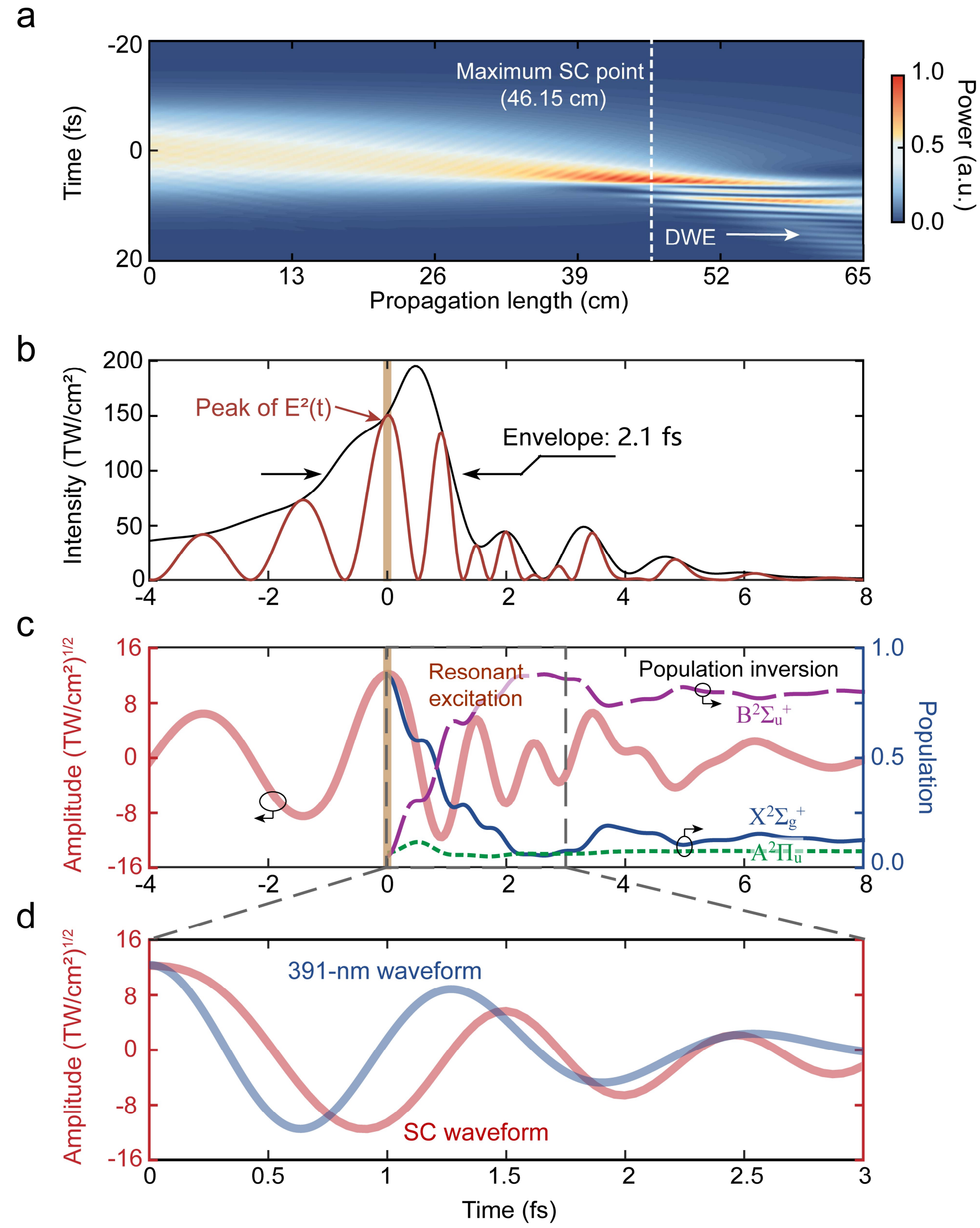

**Fig. 2 | Soliton self-compression in the tapered HCF and population dynamics of $N_2^+$ in the sub-cycle, self-compressed pulse. a** Simulated temporal evolution of an 800 nm, 12 fs, 35 μJ pulse in a 65 cm-long $N_2$-filled tapered HCF at the gas pressure of 450 mbar, corresponding to the experimental condition in Fig. 1c. SC, self compression; DWE, dispersive wave emission. **b** Pulse envelope and electric field square at the maximum SC point. **c** Time-dependent populations in the three states, within the electric field in **b** and at an alignment angle of $N_2$ of 45°. **d** Comparison of the SC waveform in (**c**) and the waveform of a 2.1 fs, 391-nm pulse.

As shown in Fig. 2c, population inversion between the $B^2\Sigma_u^+$ and $X^2\Sigma_g^+$ in the self-compressed pulse is established rapidly within ~2 fs. Next, owing to self-steepening, the trailing edge of the pulse drops sharply, freezing the inversion in place. The final population of the $X^2\Sigma_g^+$ state is ~0.125, almost entirely in the vibrational ground state ($v = 0$), while that of the $B^2\Sigma_u^+$ state is ~0.8, with 0.55 in the vibrational ground state ($v' = 0$) and 0.22 in the first excited vibrational state ($v' = 1$). Both vibrational levels are inverted with respect to the $X^2\Sigma_g^+$ ($v = 0$) state (see Supplementary Fig. S3), accounting for the 391-nm and 358-nm spectral lines observed in experiment.

This pronounced population inversion can be meanly attributed to the single-photon-induced resonant coupling between the $X^2\Sigma_g^+$ and $B^2\Sigma_u^+$ states, as evidenced by the close match between the trailing-edge waveform of the self compressed pulse and the 391-nm waveform (see Fig. 2d). To confirm this picture, we simulated the population dynamics driven by 2.1-fs, 150 TW/cm$^2$ Gaussian-shaped pulses centered at 391 nm and 800 nm, respectively. The 391-nm pulse reproduces a population inversion similar to that in Fig. 2c, whereas the 800-nm pulse does not, see Supplementary Fig. S4 for more details.

It should be noted that the process shown in Fig. 2c is in marked contrast to the multi-cycle pumping regime, where the populations in $X^2\Sigma_g^+$ and $B^2\Sigma_u^+$ oscillate periodically with the trailing electric field, and the establishment of a stable population inversion between these two states realise greatly on $X^2\Sigma_g^+$-$A^2\Pi_u$ coupling[12, 13, 41-43]. For example, we simulated the population dynamics driven by 12 fs, 800 nm Gaussian-shaped pulses, and reproduced phenomena consistent with previous studies[12, 13], such as the pronounced $X^2\Sigma_g^+$-$A^2\Pi_u$ coupling, see Supplementary Fig. S5. Notably, for 10-40 fs pump pulses, achieving comparable population inversion requires raising the peak intensity beyond 200 TW/cm$^2$, corresponding to an increase in pump energy by more than one orders of magnitude. We therefore attribute the high-efficiency and low-threshold $N_2^+$ lasing observed in our experiment primarily to the resonantly-excited population inversion enabled by the asymmetric self-compressed sub-cycle pulse.

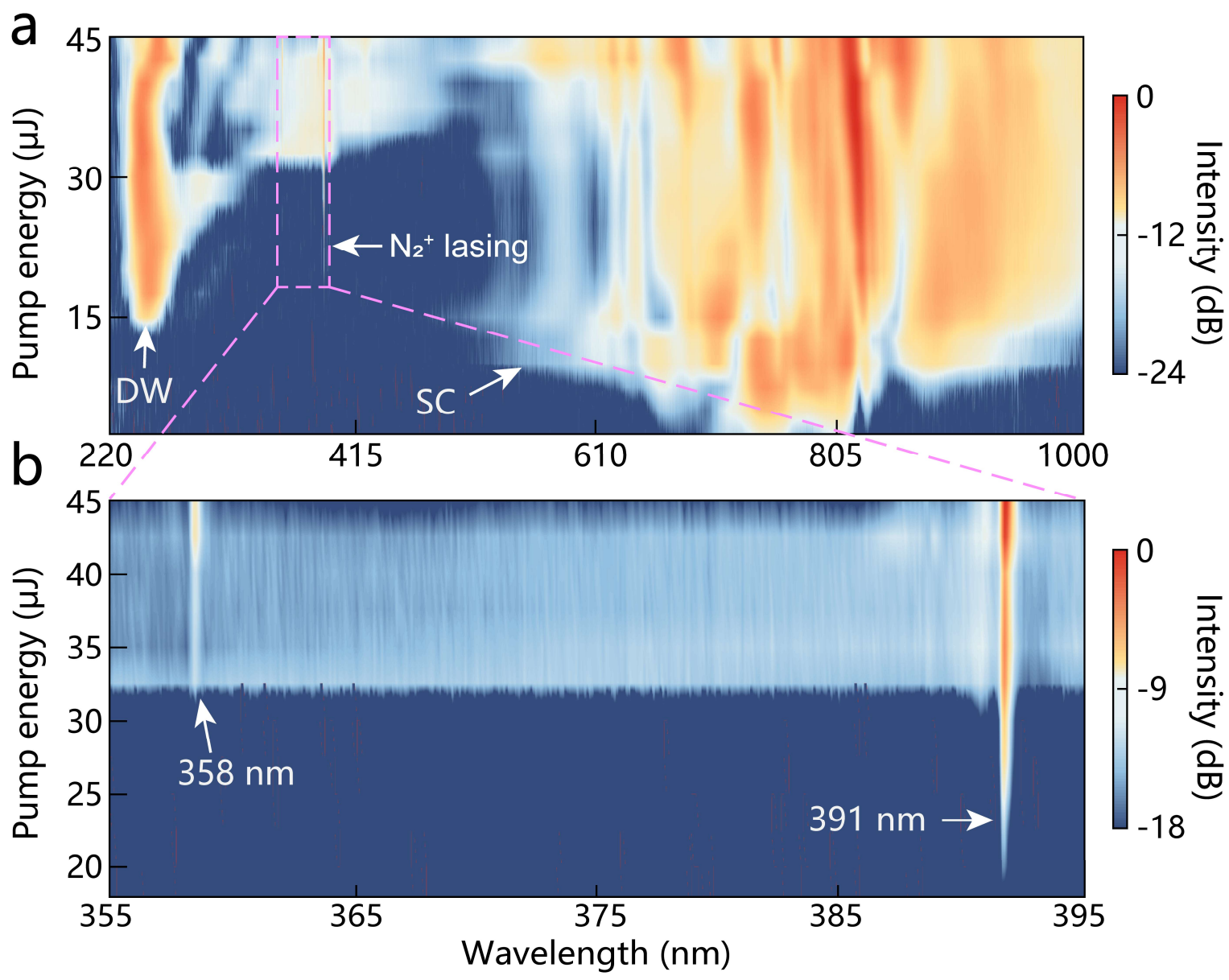

**Fig. 3 | Measured output spectra from the tapered HCF at different pump energies. a** Measured output spectral evolution from the tapered HCF as a function of pump energy. The system parameters are the same as that Fig. 1c. DW, dispersive wave; SC, self compression. **b** The spectral evolution of $N_2^+$ lasing signals magnified from (**a**).

The picture of sub-cycle soliton dynamics driving resonant $X^2\Sigma_g^+$ - $B^2\Sigma_u^+$ coupling can fully explain our experimental observations. Figure 3 shows the measured output spectra from the tapered HCF as a function of pump energy. For pump energies below 17 μJ, the spectral evolution is dominated by soliton dynamics: sub-cycle self-compression drives multi-octave spectral broadening and generates dispersive waves at ~245 nm. A weak 391-nm signal emerges once the pump energy exceeds 17 μJ, at which point the peak intensity in the fibre is estimated to be ~120 TW/cm$^2$, which is much lower than the typical intensity required in conventional multi-cycle pumping schemes, highlighting the crucial role of the asymmetric self-compressed waveform in producing population inversion. When the pump energy is increased beyond 30 μJ, the 358-nm $N_2^+$ lasing line appears and the 391-nm energy grows rapidly, as shown in Fig. 3b and Supplementary Fig. S6. A broadband pedestal spanning 300-500 nm also emerges; this pedestal originates from self-steepening effect[4], corresponds to the trailing-edge waveform in Fig. 2d, and serves to resonantly enhance the population transfer from $X^2\Sigma_g^+$ to $B^2\Sigma_u^+$.

To examine the maximum achievable $N_2^+$-lasing energy, we measured the UV spectrum directly at the HCF output port under three pump energies of 20, 35, and 50 μJ. As shown in Fig. 4, the highest 391-nm energy exceeds 160 nJ at 50 μJ pump energy; however, the beam profile of the $N_2^+$-lasing signals at this pump level exhibits slight distortion compared with those at the 20 and 35 μJ cases, which we attribute to weak ionization at the fibre entrance that perturbs the launched mode pattern.

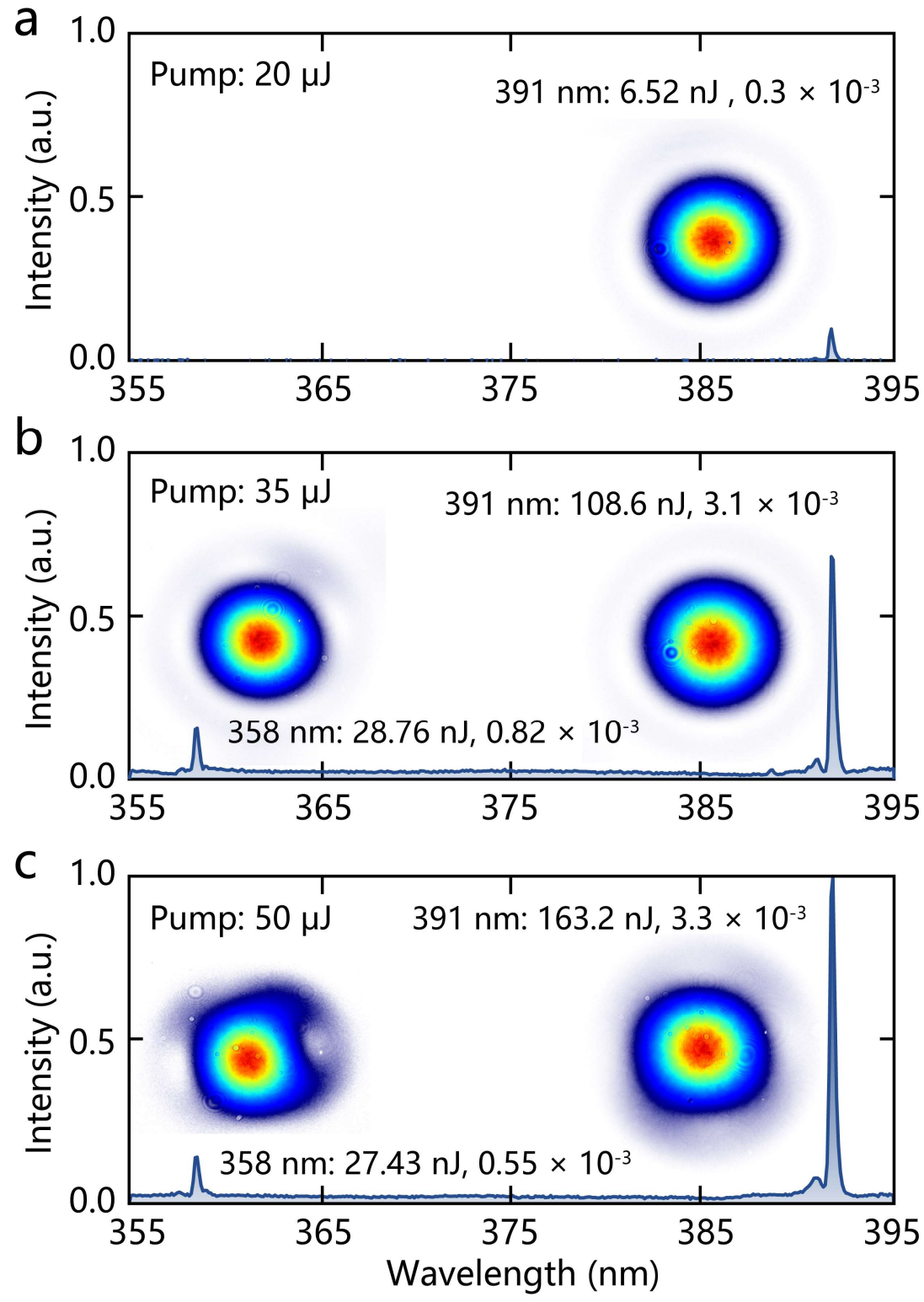

**Fig. 4 | $N_2^+$ lasing signals and beam profiles at different pump energies. a** 20 μJ pump, **b** 35 μJ pump and **c** 50 μJ pump. The spectra and the beam profiles are measured directly at the HCF output port.

The experimentally generated 391-nm $N_2^+$-lasing signal reaches 163.2 nJ pulse energy with a maximum conversion efficiency of $3.3 \times 10^{-3}$ at a pump energy of only 50 μJ. As shown in Fig. 5, this represents a dramatic improvement in both efficiency and threshold over conventional air-lasing sources[17, 35-37, 44, 45], which confirms the advantage of the HCF platform in combining soliton self-compression with sub-cycle-field-induced population inversion. The pulse energy of 391-nm $N_2^+$-lasing signal shown here has not yet reached the microjoule level, a problem that can be addressed by utilizing the scaling law of soliton dynamics: the intensity of the signal can increase linearly with the increase of pump energy, which requires proportionally extending the HCF length and core area, and reducing the filling pressure in the same proportion[3].

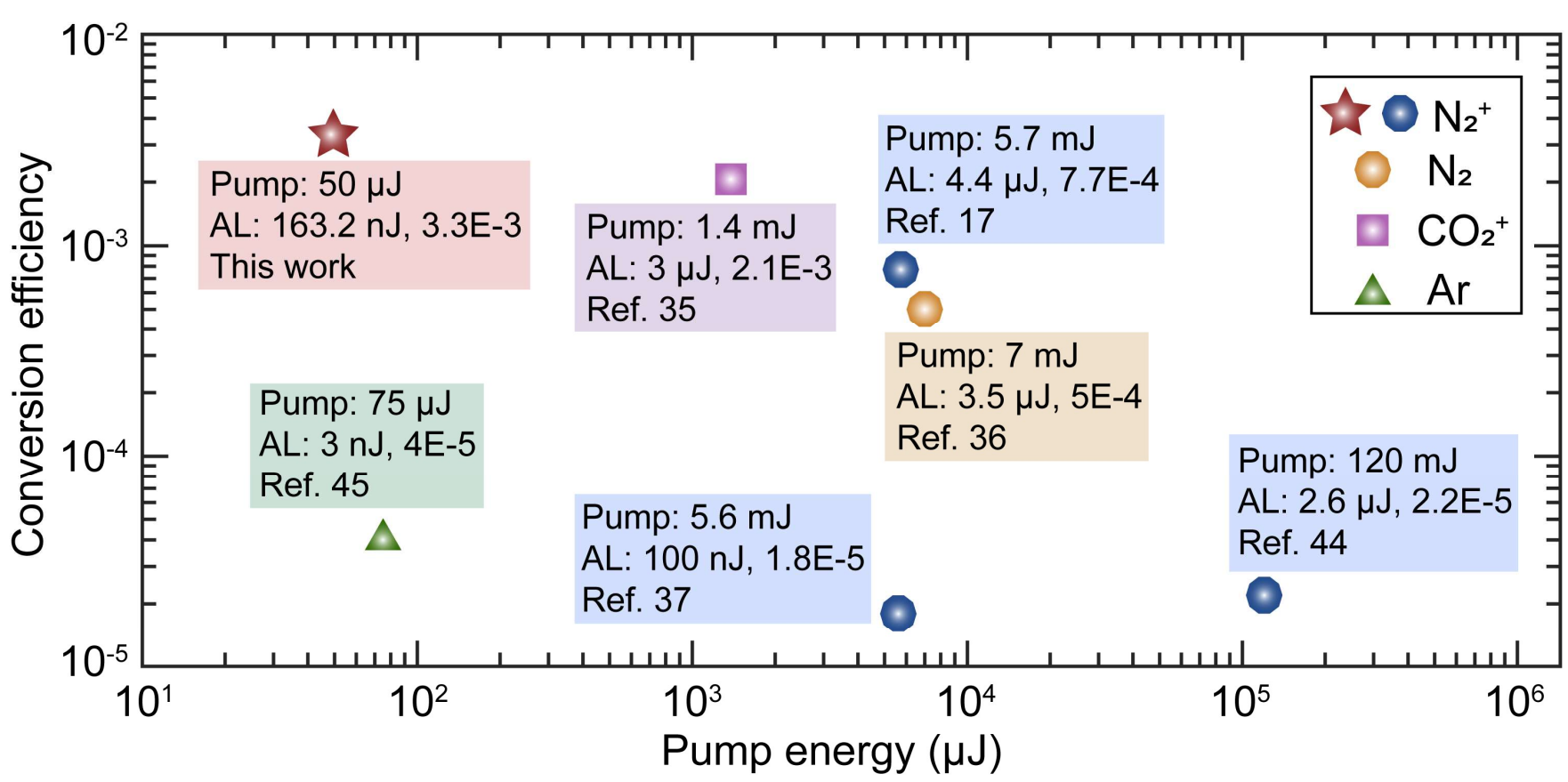

**Fig. 5 | Comparison with state-of-the-art air lasing (AL) sources driven by ultrafast lasers.** The parameters marked by different shapes and colors represent different AL type, as shown in the legends.

Beyond providing narrowband $N_2^+$-lasing signal, the soliton dynamics can also deliver broadly tunable UV DW pulses. In the experiment, we varied the $N_2$ gas pressure in the tapered HCF from 300 mbar to 800 mbar, corresponding to a tuning of the phase-matching wavelength of DW emission from ~220 nm to ~280 nm[46]. We also adjusted accordingly the pump pulse energy so as to achieve clear $N_2^+$-lasing signal at each parameter, and observed continuous central-wavelength tuning (ranging from 220 nm to 280 nm) of the DW emission, meanwhile the strongest $N_2^+$-lasing signal remained at 391 nm, see Supplementary Fig. S7. These DW pulses, with microjoule-level pulse energy and sub-10-fs pulse width[4, 47, 48], are naturally collinear with the 391-nm $N_2^+$-lasing signals. Together, they can form a compact and tunable source for high-resolution pump-probe spectroscopy[14-17].

In addition to providing advanced light sources, our results build a bridge between sub-cycle soliton dynamics and strong-field-driven gas lasing. We have shown that, inside the HCF, the self-compressed pulse can approach ~2 fs, confining light-matter interaction to the sub-two-cycle scale and granting access to new physical process. For instance, recent studies have shown that near-single-cycle drivers render the lasing signal highly sensitive to the carrier envelope phase (CEP) of the pump pulse, enabling CEP-controlled lineshape tailoring of broadband UV radiation[34, 35, 49] in $N_2$ and $CO_2$. Our scheme, which employs relatively accessible few-cycle pulses and takes advantage of nonlinear self-compression in a HCF, provides a versatile and practical route to such sub-cycle-scale investigations.

In conclusion, we have shown that soliton dynamics in a tapered HCF can decouple the $N_2^+$ lasing from femtosecond filamentation, largely improving conversion efficiency, pump threshold and beam quality of 391-nm $N_2^+$-lasing signal. From the application viewpoint, the demonstrated narrow band $N_2^+$-lasing source, together with the broadly tunable ultrafast UV DW pulses, forms an efficient and compact laser set-up that may render great application potential in combustion diagnosis and environmental monitoring. From a physics perspective, our work establishes a bridge between the two largely separated fields of soliton dynamics and strong-field-induced gas lasers, offering a powerful new tool for probing molecular dynamics at sub-cycle timescales.

## Methods

**Experimental set-up.** The high-efficiency 391-nm $N_2^+$-lasing system consists of two stages, including a HCF compression stage and a soliton stage, driven by a commercial Ti:sapphire femtosecond laser, delivering 800 nm, 40 fs pulses at a repetition rate of 1 kHz. In the first stage, the pump pulses with an energy of 2.2 mJ were compressed using an HCF compression set-up, consisting of a 1-m-long capillary fibre, with an inner diameter of 500 µm, filled with Ne gas, followed by a broadband attenuator and two pairs of chirped mirrors (DCMP175, UMC10-15FS, Thorlabs). The system can compress the input pulses from 40 fs to 12 fs at 1.2 bar Ne-gas pressure (see Supplementary note 1 for more details). The transmission efficiency of this system is ~60%, corresponding to a maximum output pulse energy of 1.3 mJ. In the second stage, a $N_2$-filled tapered capillary fibre (core diameter decreases linearly from 160 to 96 µm in 65 cm length) was used for high-efficient $N_2^+$ lasing. See the Supplementary Fig. S1 for more details of the set-up.

**Pulse spectra and energy measurements.** Pulse spectra were measured using two spectrometers: one for the UV range (Maya2000-Pro, 200-400 nm, Ocean Insights) and one for the VIS to NIR range (Maya2000-Pro, 210-1100 nm, Ocean Optics). Both were calibrated. The full optical spectrum at the HCF output (Figs. 1c and 3a) was obtained by combining data from both spectrometers and interpolating across the overlapping region. For these measurements, the output beam was split by a fused silica wedge. The transmitted beam was directed into an integrating sphere coupled to the UV spectrometer, and the reflected beam into another integrating sphere coupled to the VIS-NIR spectrometer.

Input pulse energy was measured with a thermal power meter (3A-P, Ophir). The energies of DW and $N_2^+$-lasing signals were estimated by integrating the measured spectral intensity over the respective wavelength ranges. The spectral sensitivity of the UV spectrometer was calibrated following the procedure described in our previous work[4, 7, 26, 40, 50]. Briefly, a broadband UV supercontinuum spanning 200-400 nm was generated, and a series of narrow bandpass filters (Pelham Research Optical for 200-260 nm with 20 nm interval; FGUV5 and FGUV11, Thorlabs for 240-395 nm) were used to select specific wavelength intervals. The energy of each filtered portion was measured with a photodiode (PD300-UV, Ophir), and the corresponding spectral intensity was recorded with the UV spectrometer, yielding a wavelength-dependent calibration factor between the spectrometer signal and the spectral energy density.

**Beam profile measurements.** Near-field beam profiles of the output 358- and 391-nm $N_2^+$-lasing signals (Figs. 1d and 4a-c) were measured using optical filters (FBH390-10 and FBH360-10, Thorlabs) and a CCD camera (BGS-USB3-SP932U, Ophir-Spiricon).

## Data availability

The data generated in this study is available from the corresponding author upon reasonable request.

## Code availability

The code used in this paper is available from the corresponding author upon reasonable request.

## Acknowledgements

The authors would like to thank Prof. J. Yao for insightful discussion. This work was supported in part by the National Natural Science Foundation of China (No. W2541021 to M.P., No. 62505330 to J.Y.P., No. 12388102 to R.X.L.), the Strategic Priority Research Program of the Chinese Academy of Science (No. XDB0650000 to Z.Y.H. and M.P.), the Shanghai Science and Technology Plan Project Funding (No. 23JC1410100 to Z.Y.H. and M.P.), the National Postdoctoral Program for Innovative Talents (No. BX20250361 to J.Y.P.), the China Postdoctoral Science Foundation (No. 2025M780803 to J.Y.P.), Shanghai Municipal Science and Technology Major Project (Z.Y.H. and M.P.), Fuyang High-level Talent Group Project (J.P.H, W.B.H., X.J. and M.P.).

## Author contributions

Z.Y.H., M.P. and Y.X.L. conceived the work. T.D.C, P.T.W and Z.Y.H carried out the experiments. W.B.H., J.P.H., X.J., Y.X.L. and R.X.L. provided necessary experimental equipment and resources. T.D.C. performed the numerical simulations. T.D.C., Z.Y.H., H.L.X., Y.L., M.P. and Y.X.L. made the theoretical and experimental analysis. T.D.C., Z.Y.H., M.P. and Y.X.L. wrote the manuscript, Z.Y.H., M.P. and Y.X.L. supervised the project. All authors contributed to the discussion of the results and the editing of the manuscript.

## Competing interests

The authors declare no competing interests.

## Additional information

Supplementary information is available for this paper at xxxxxx.

Supplementary Information for

# "Low-threshold efficient $N_2^+$ lasing driven by sub-cycle soliton dynamics in a hollow waveguide"

Tiandao Chen[1], Zhiyuan Huang[1,2,3,*], Jinyu Pan[1], Donghan Liu[1,2,3], Pengtao Wang[2,3], Xinglin Zeng[2,3], Jinxin Zhan[2,3], Jiapeng Huang[2,3], Wenbin He[2,3], Xin Jiang[2,3], Huailiang Xu[4], Yi Liu[5], Meng Pang[1,2,3,6,*], Yuxin Leng[1,6,*] and Ruxin Li[1,7]

[1]State Key Laboratory of Ultra-intense Laser Science and Technology, Shanghai Institute of Optics and Fine Mechanics (SIOM), Chinese Academy of Sciences (CAS), Shanghai 201800, China

[2]Russell Centre for Advanced Lightwave Science, Shanghai Institute of Optics and Fine Mechanics (SIOM) and Hangzhou Institute of Optics and Fine Mechanics (HIOM), Hangzhou 311400, China

[3]Zhejiang Key Laboratory of Microstructured Specialty Optical Fiber, Hangzhou Institute of Optics and Fine Mechanics (HIOM), Hangzhou 311400, China

[4]School of Optoelectronic Engineering, Xidian University, Xi'an 710071, China

[5]School of Optical-Electrical and Computer Engineering, University of Shanghai for Science and Technology, Shanghai 200093, China

[6]Hangzhou Institute for Advanced Study, University of Chinese Academy of Sciences, Hangzhou 310024, China

[7]Zhangjiang Laboratory, Shanghai 201210, China

[*]Corresponding author: huangzhiyuan@siom.ac.cn; pangmeng@siom.ac.cn; lengyuxin@siom.ac.cn

## Supplementary note 1: Experimental set-up and pulse diagnosis

As shown in Fig. S1a, the experimental set-up for the low-threshold efficient $N_2^+$ lasing system includes two stages: hollow capillary fibre (HCF) compression and soliton driven $N_2^+$ lasing. This system is driven by a commercial Ti:sapphire femtosecond laser (Solstice Ace 80L-35F-1K-HP-T, Spectra-Physics), producing 800 nm, 2.2 mJ, 40 fs pulses with a repetition rate of 1 kHz. The whole experimental system is installed on an optical table, and the physical size of this system is 2 m × 1 m. In the first stage, the 800 nm pulses from the driving laser were first launched into the Ne-filled HCF using a coated plano-convex lens with a focal length of 2 m. The input pulse energy was controlled by a broadband attenuator based on a half-wave plate (HWP) and a wire grid polarizer (WGP). The HCF has a core diameter of 500 μm and a length of 1 m. The input and output ends of the HCF were sealed with 0.5-mm-thick coated fused silica (FS) windows. The laser pulses experienced a spectral broadening in the Ne-filled HCF due to self-phase modulation. The pulse energy at the output port of the HCF was measured to be ~1.8 mJ, corresponding to a transmission efficiency of ~82%. The output pulses were collimated by a coated concave mirror with a focal length of 1 m, and the pulse energy was controlled by another broadband attenuator, which is composed by a HWP and a silica plate, as shown in Fig. S1a. Two pairs of chirped mirrors (DCMP175, UMC10-15FS, Thorlabs) were used to compensate for the output pulses. When the HCF was filled with 1.2 bar Ne gas, the laser pulses could be compressed from 40 fs to 12 fs, measured using a home-built second-harmonic generation frequency-resolved optical gating (SHG-FROG) set-up, see Figs. S1b-d. The efficiency of the HCF compression stage is ~60%, corresponding to a maximum pulse energy of 1.3 mJ.

In the second stage, the compressed pulses from the first stage were coupled into another N2-filled tapered HCF with a core diameter decreasing linearly from 160 to 96 μm and a length of 65 cm, using a lens with a focal length of 30 cm. The input and output ports of the second HCF were sealed with a 0.5-mm-thick coated FS window and a 0.5-mm-thick uncoated FS window, respectively. The efficiency of the soliton-driven $N_2^+$ lasing stage is ~20%, including ~60% coupling efficiency and ~34% transmission efficiency. The $N_2^+$ lasing signals generated at the output of the second HCF will be diagnosed, including spectral, energy and beam profile measurements.

We further examined the spatial coherence of the 391- and 358-nm signals using the setup shown in Fig. S2a. The narrowband signals were divided into two parts by the first D-shaped mirror (DM1). These two pulses were focused into a CCD camera using a concave mirror with a focal length of 10 cm. As shown in Fig. S2b, both 391-nm signal and 358-nm signals can produce clear interference pattern, which indicates that these signals have good spatial coherence and are originate from $N_2^+$ lasing rather than spontaneous emission. The degree of spatial coherence of 391 nm and 358 nm signals are estimated to be 0.72 and 0.79, respectively. These estimated values could merely give the lower limit values of the coherence degree, and the measurement accuracy is largely limited by the spatial resolution of the CCD camera used in the experiment.

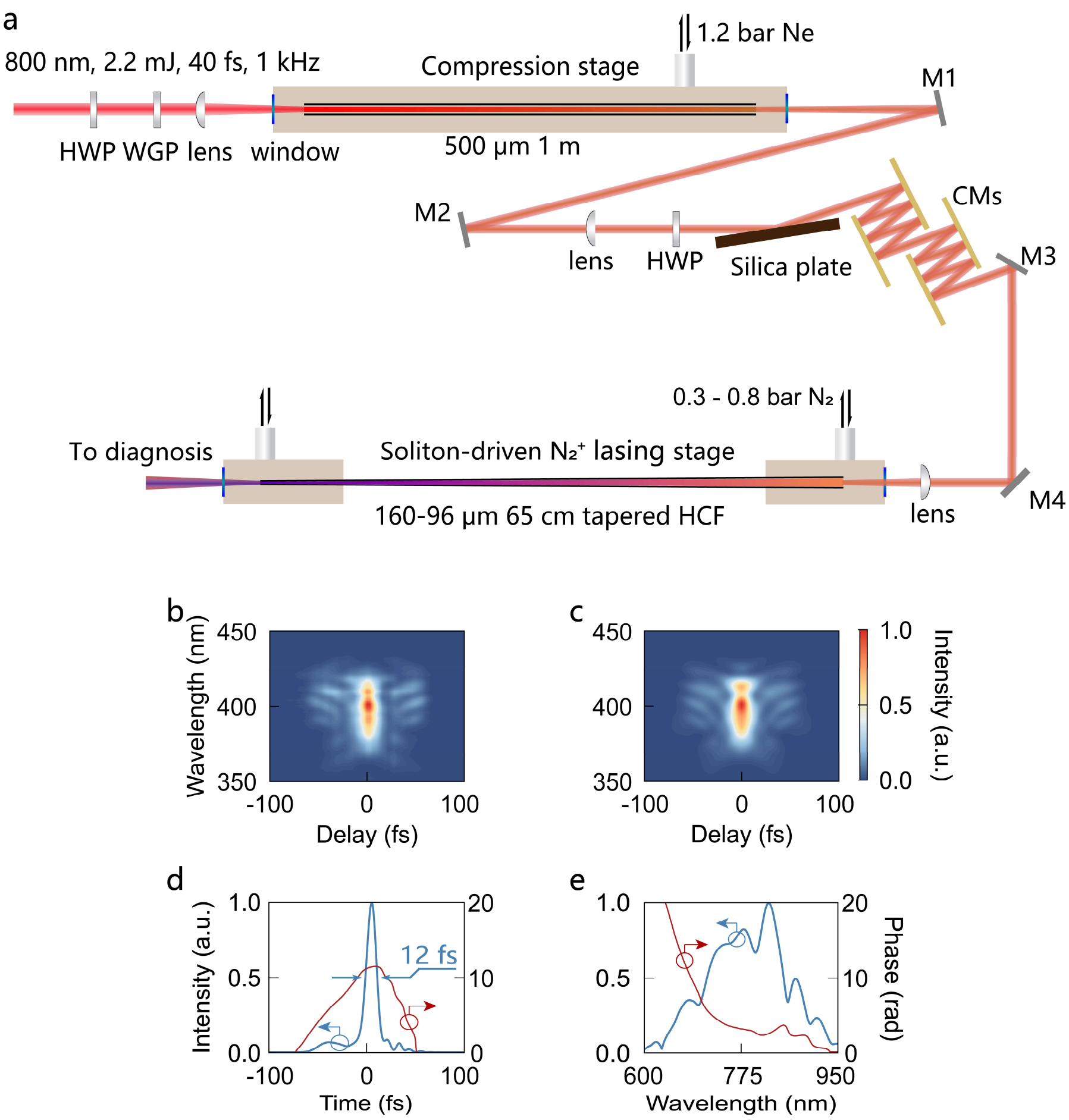

**Supplementary Fig. S1. Experimental set-up of $N_2^+$ lasing generation and pump pulse characterization. a** HWP, half-wave plate; WGP, wire grid polarizer; CMs, chirped mirrors. **b** Measured FROG trace of the pump pulse. **c** Retrieved FROG traces, with a retrieving error of 0.6%. **d** Retrieved temporal (blue line) and phase (red line) profiles. **e** Retrieved spectral (blue line) and phase (red line) profiles.

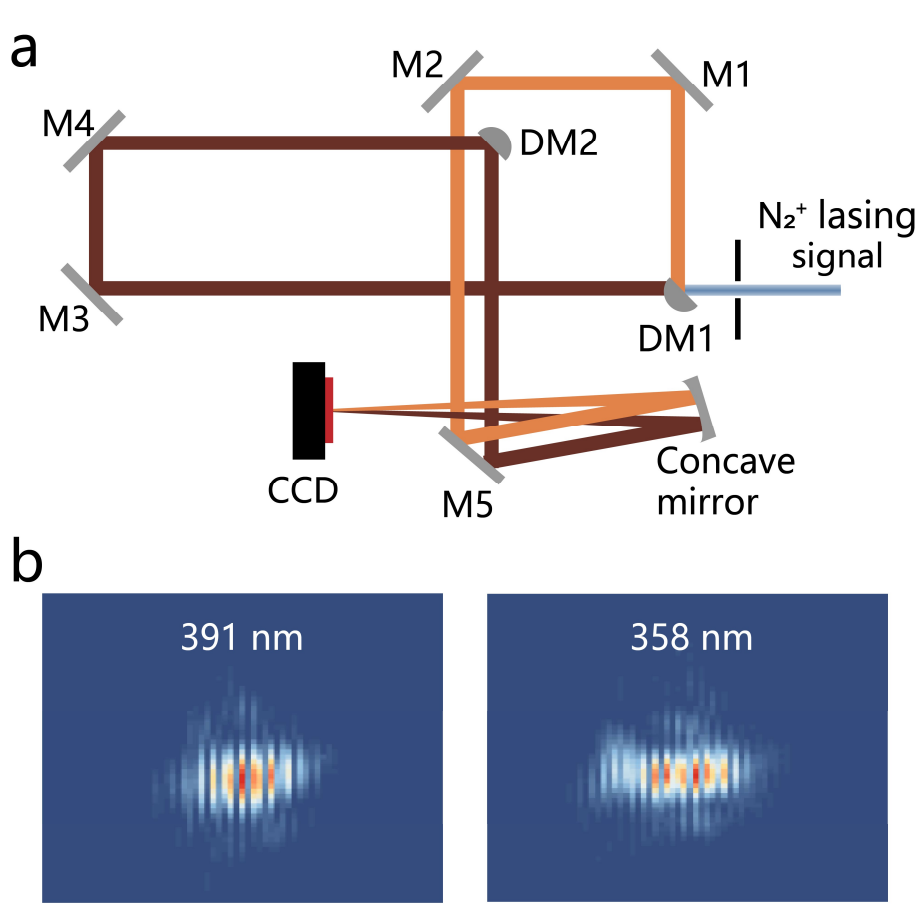

**Supplementary Fig. S2. Spatial coherence of the $N_2^+$ lasing signal. a** DM1 and DM2, D-shaped mirrors; CCD, Charge coupled Device. **b** Measured interference pattern at 391 nm and 358 nm using setup in (**a**).

## Supplementary note 2: Numerical simulation model

### 2.1 Simulation model of sub-cycle soliton self-compression

The soliton self-compression of ultrafast laser pulses in gas-filled tapered HCF can be simulated using the single-mode unidirectional pulse propagation equation (UPPE), which can be expressed as[1-5]:

$$\frac{\partial \tilde{E}(z,\omega)}{\partial z} = i\left(\beta(z,\omega) - \frac{\omega}{v_g(z)}\right)\tilde{E}(z,\omega) - \frac{\alpha(z,\omega)}{2}\tilde{E}(z,\omega) + i\frac{\omega^2 \tilde{P}_{NL}(z,\omega)}{2c^2\varepsilon_0\beta(z,\omega)} \tag{S1}$$

where $\tilde{E}$ is the electric field in frequency domain, $\omega$ is the angular frequency, $z$ is the propagation distance along the tapered HCF, $\beta$ is the propagation constant, $v_g = (\partial\beta/\partial\omega|_{\omega_0})^{-1}$ is the group velocity at the pump central frequency $\omega_0$, and $\alpha$ the HCF loss. $c$ is the light speed in vacuum, $\varepsilon_0$ is the vacuum permittivity, and $\tilde{P}_{NL}$ is nonlinear polarization in the frequency domain. The nonlinear polarization includes the Kerr effect, ionization-induced plasma polarization and Raman effect, which can be given as[4]:

$$\tilde{P}_{NL}(z,\omega) = F\left[\varepsilon_0\chi^{(3)}E(z,t)^3 + P_{ion}(z,t) + P_{Raman}(z,t)\right] \tag{S2}$$

where $F$ represents the Fourier transform, $\chi^{(3)}$ is the third-order nonlinear susceptibility that is related to the Kerr effect, $E$ is the electric field in the time domain, $t$ is the time in a reference frame travelling with a group velocity. $P_{ion}$ is the ionization-induced plasma polarization given by[2-5]:

$$\frac{\partial P_{ion}(z,t)}{\partial t} = \frac{I_P}{E(z,t)}\frac{\partial \rho(z,t)}{\partial t} + \frac{e^2}{m_e}\int_{-\infty}^{t}\rho(z,t')E(z,t')dt' \tag{S3}$$

where $I_P$ is the ionization potential of the gas, $\rho$ is the plasma density, $e$ is the electronic charge, and $m_e$ is the mass of an electron. The plasma density can be described as:

$$\frac{\partial \rho}{\partial t} = W(I)(\rho_{nt} - \rho) + \frac{s}{I_P}\rho I \tag{S4}$$

where $W$ is the optical-field-induced ionization rate which depends on the intensity of laser pulses, $I$ is the laser intensity of the pulses, $\rho_{nt}$ is the neutral gas density, and $s$ is the cross section describing the process of the collisional ionization. Since the recombination time of the $N_2$ is on the nanosecond scale, the electron recombination term is not included in Eq. (S4).

The Raman effect term includes vibrational response and rotational response $P_{Raman}(z,t) = P_{vib}(z,t) + P_{rot}(z,t)$, where $P_{vib}(z,t)$ is given by[4, 6]:

$$P_{vib}(z,t) = -\frac{2\rho_{nt}\varepsilon_0^2\pi}{\mu v_{vib}}\left(\frac{\partial \alpha_m}{\partial Q_m}\right)^2 E(z,t)\int_0^{\infty} h(v_{vib}, \Delta v_{vib}, t')E(z,t-t')^2 dt' \tag{S5}$$

where $\mu$ is the reduced molecular mass, $\nu_{vib}$ is the vibration frequency, $\alpha_m$ is the isotropic averaged molecular polarizability and $Q_m$ is the ensemble-averaged molecular stretch, and $\Delta\nu_{vib}$ the full-width at half maximum (FWHM) of the Raman transient of molecular vibration. $h(\nu,\Delta\nu,t)$ is the Raman response function given by $h(\nu,\Delta\nu,t)=\sin(2\pi\nu t)\exp(-\pi\Delta\nu t)$. Rotational response $P_{rot}$ is given by[4, 6]:

$$P_{rot}(z,t)=\frac{(4\pi)^3\rho_{nt}\varepsilon_0{}^2\Delta\alpha_m{}^2}{15h}E(z,t)\times \\ \sum_{J=0}^{50}\frac{(J+1)(J+2)}{(2J+3)}\left(\frac{\rho_{J+2}}{2J+5}-\frac{\rho_J}{2J+1}\right)\int_0^\infty h(\nu_{rot}^J,\Delta\nu_{rot}^J,t')E(z,t-t')^2dt' \tag{S6}$$

where $\Delta\alpha_m$ is the molecular polarizability anisotropy, $h$ is the Planck constant, $J$ is the rotational quantum number, $\rho_J$ the thermal population at rotational level $J$, $\nu_{rot}^J$ the rotation frequency at rotational level $J$, and $\Delta\nu_{rot}^J$ the FWHM of Raman transient at rotational level $J$. Up to 50 rotational levels are considered in the simulation.

The core radius of the tapered HCF $a$ decreases linearly with $z$, and the corresponding propagation constant can be written as[7]:

$$\beta(z,\omega)=\frac{\omega}{c}\sqrt{n_{gas}^2(z,\omega)-\frac{u_{11}^2c^2}{\omega^2a(z)^2}} \tag{S7}$$

where $n_{gas}$ is the refractive index of the filling gas, which depends on light frequency, gas pressure, and temperature. $u_{11}$ is the first zero of zeroth-order Bessel function, which is ~2.4048. The linear loss of the HCF can be given as[7]:

$$\alpha(\omega)=\frac{u_{11}^2c^2}{\omega^2a(z)^3}\frac{\eta^2+1}{\sqrt{\eta^2-1}} \tag{S8}$$

where $\eta$ is the ratio of the refractive index of the cladding medium to that of the filling gas.

In the simulation, we used 12-fs (full width at half maximum), 35-μJ, 800-nm, Gaussian-shape pulses as input. The numerically results include electric fields at different HCF positions, and was extracted to simulate the population dynamics in $N_2^+$.

### 2.2 Simulation model of field-induced population dynamics in the $N_2^+$

The population dynamics of $X^2\Sigma_g^+$, $A^2\Pi_u$ and $B^2\Sigma_u^+$ electronic states in a given pulse can be simulated using the time-dependent Schrödinger equation[8, 9]:

$$i\hbar \frac{\partial \mathbf{\Psi}(r,t)}{\partial t} = \left( -\frac{\hbar^2}{2\mu}\frac{\partial^2}{\partial r^2} + \mathbf{V}(r) + E(t)\cos\theta \mathbf{D}_{\mathrm{XB}}(r) + E(t)\sin\theta \mathbf{D}_{\mathrm{XA}}(r) \right)\mathbf{\Psi}(r,t) \qquad \text{(S9)}$$

where $\mathbf{\Psi} = (\psi_X, \psi_A, \psi_B)^T$ is a column vector containing the nuclear wave functions at $X^2\Sigma_g^+$, $A^2\Pi_u$ and $B^2\Sigma_u^+$ electronic states, $r$ is the internuclear distance, $\mu$ is the reduced mass of $N_2^+$, $\theta$ the angle between the laser polarization direction and the molecular axis, which is set to be 45°[8, 9], and $E$ is the self-compressed or Gaussian-shaped pulse with selected carrier wavelength. $\mathbf{V}$ is a 3×3 diagonal matrix, with matrix elements being the potential curves of the three electronic states[10, 11]. $\mathbf{D}_{\mathrm{XB}}$ and $\mathbf{D}_{\mathrm{XA}}$ are 3×3 dipole transition matrices[10, 11] with the only non-vanishing matrix elements being $(\mathbf{D}_{\mathrm{XB}})_{13} = (\mathbf{D}_{\mathrm{XB}})_{31} = d_{13}(r)$ and $(\mathbf{D}_{\mathrm{XA}})_{12} = (\mathbf{D}_{\mathrm{XA}})_{21} = d_{12}(r)$.

The initial nuclear wave function at ionization moment ($t = 0$) is considered to be the vibrational ground state of the electronic ground state potential curve of the neutral $N_2$ molecule[8, 12], and the initial populations at the peak of a given pulse is calculated using the molecular tunneling ionization model[13, 14]. The population in a given state (for example, $X^2\Sigma_g^+$) is defined as $p_X(t) = \int dr |\psi_X|^2$, and the population in the three states satisfies $p_X(t) + p_A(t) + p_B(t) = 1$.The population in a given the vibrational state is calculated as $p_X^v(t) = \left| \int \psi_X^*(r,t)\phi_X^v dr \right|^2$, where $\phi_X^v$ is the stationary wave function of $v^{\mathrm{th}}$ vibrational state of $X^2\Sigma_g^+$.

## Supplementary note 3: Population dynamics of $N_2^+$ in different driving pulses

In this note, we use Eq. (S9) to simulate the population dynamics of $N_2^+$ driven by different pulses. Figure S3 shows the population dynamics of the $X^2\Sigma_g^+$ ($v = 0$), $B^2\Sigma_u^+$ ($v' = 0$), and $B^2\Sigma_u^+$ ($v' = 1$) states driven by the self-compressed pulse, under the same simulation conditions as in Fig. 2c of the main text. The resonant excitation induced by the self-compressed waveform rapidly transfers population from $X^2\Sigma_g^+$ ($v = 0$) to both $B^2\Sigma_u^+$ ($v' = 0$) and $B^2\Sigma_u^+$ ($v' = 1$), establishing population inversion with respect to the ground state and thus enabling the two observed lasing lines at 391 nm and 358 nm.

Figures S4a,b display the population dynamics driven by a 2.1-fs, 391-nm pulse. Because the carrier wavelength of this pulse matches the $X^2\Sigma_g^+$ ($v = 0$) to $B^2\Sigma_u^+$ ($v' = 0$) transition, it resonantly excites population inversion, yielding results closely resembling those in Fig. 2c of the main text. This agreement confirms the essential role of the asymmetric self-compressed waveform in establishing population inversion. In contrast, an 800-nm pulse of the same duration and peak intensity fails to produce stable population inversion, as shown in Figs. S4c,d.

Figures S5 shows the population dynamics induced by a 12-fs, 800-nm, multi-cycle pulses. Here, achieving a clear population inversion between $B^2\Sigma_u^+$ and $X^2\Sigma_g^+$ requires raising the peak intensity beyond 200 TW/cm$^2$. Meanwhile, because the trailing field comprises multiple optical cycles, resonant coupling between $X^2\Sigma_g^+$ and $A^2\Pi_u$ sets in, and the population of the $A^2\Pi_u$ state grows almost monotonically. Ultimately, population inversion between $B^2\Sigma_u^+$ and $X^2\Sigma_g^+$ is established via field-induced three-state coupling, a behavior consistent with the results reported in Refs. [8,9].

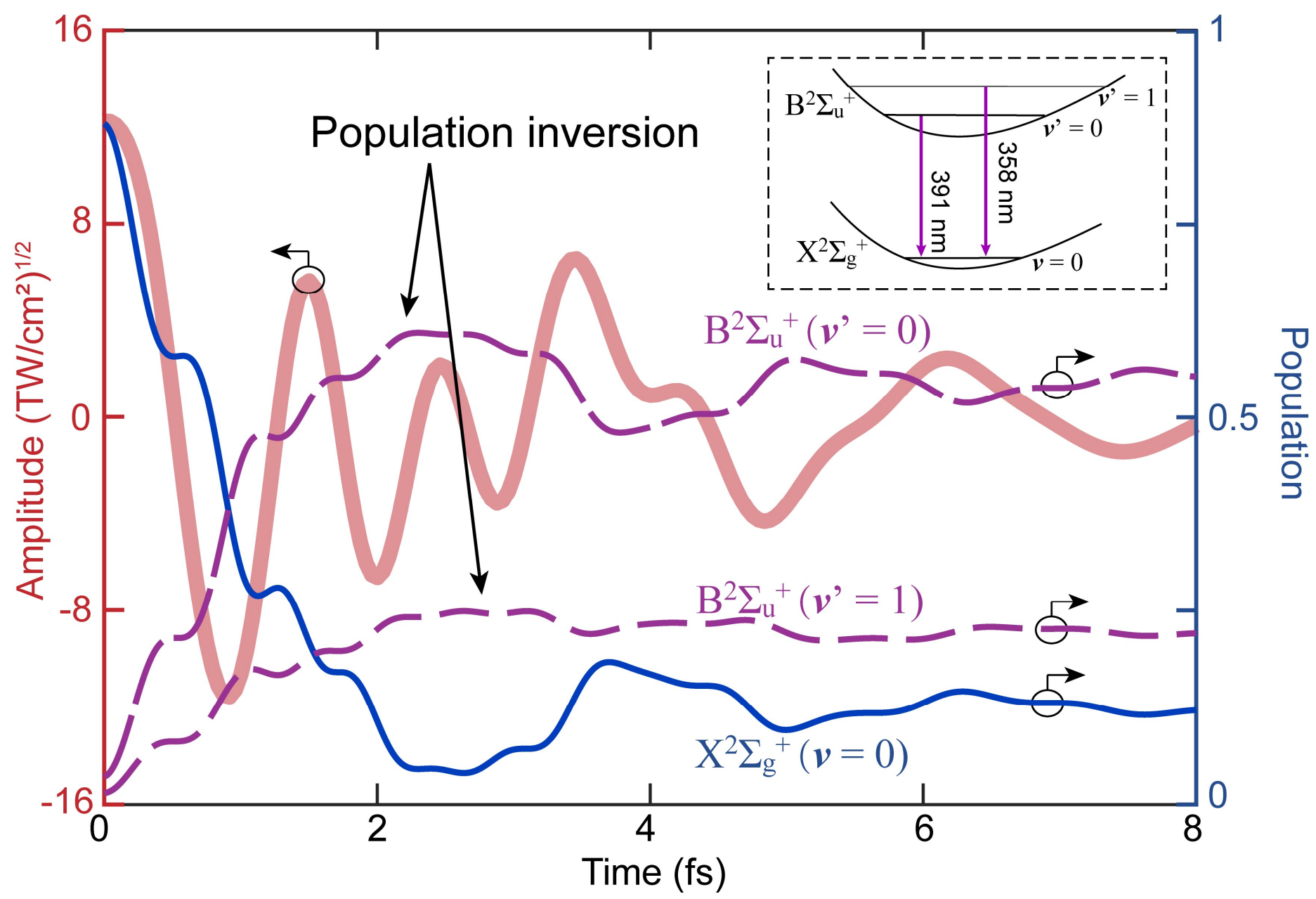

**Supplementary Fig. S3. Population dynamics in different vibrational states.** The simulation conditions are same as that in Fig. 2c of the main text. The inset shows the energy-level diagram of $N_2^+$ and the lasing wavelengths of the $B^2\Sigma_u^+$ ($v' = 0$) to $X^2\Sigma_g^+$ ($v = 0$) and $B^2\Sigma_u^+$ ($v' = 1$) to $X^2\Sigma_g^+$ ($v = 0$) transitions.

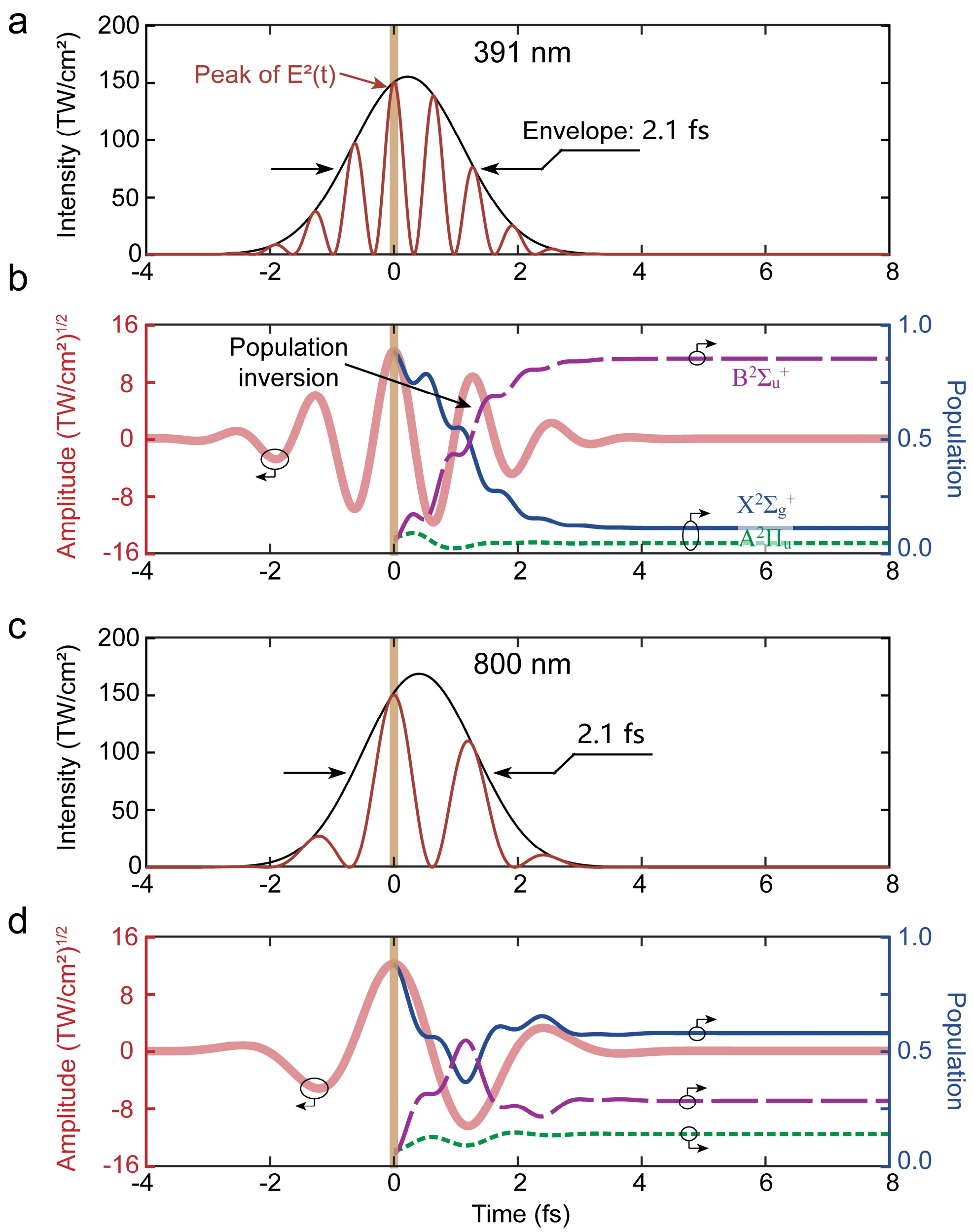

**Supplementary Fig. S4. Comparison of population dynamics of $N_2^+$ driven by 2.1-fs, 150 TW/cm$^2$, Gaussian-shaped pulses with different central wavelength**. **a,b** 391 nm pulse and corresponding population dynamics. **c,d** 800 nm pulse and corresponding population dynamics.

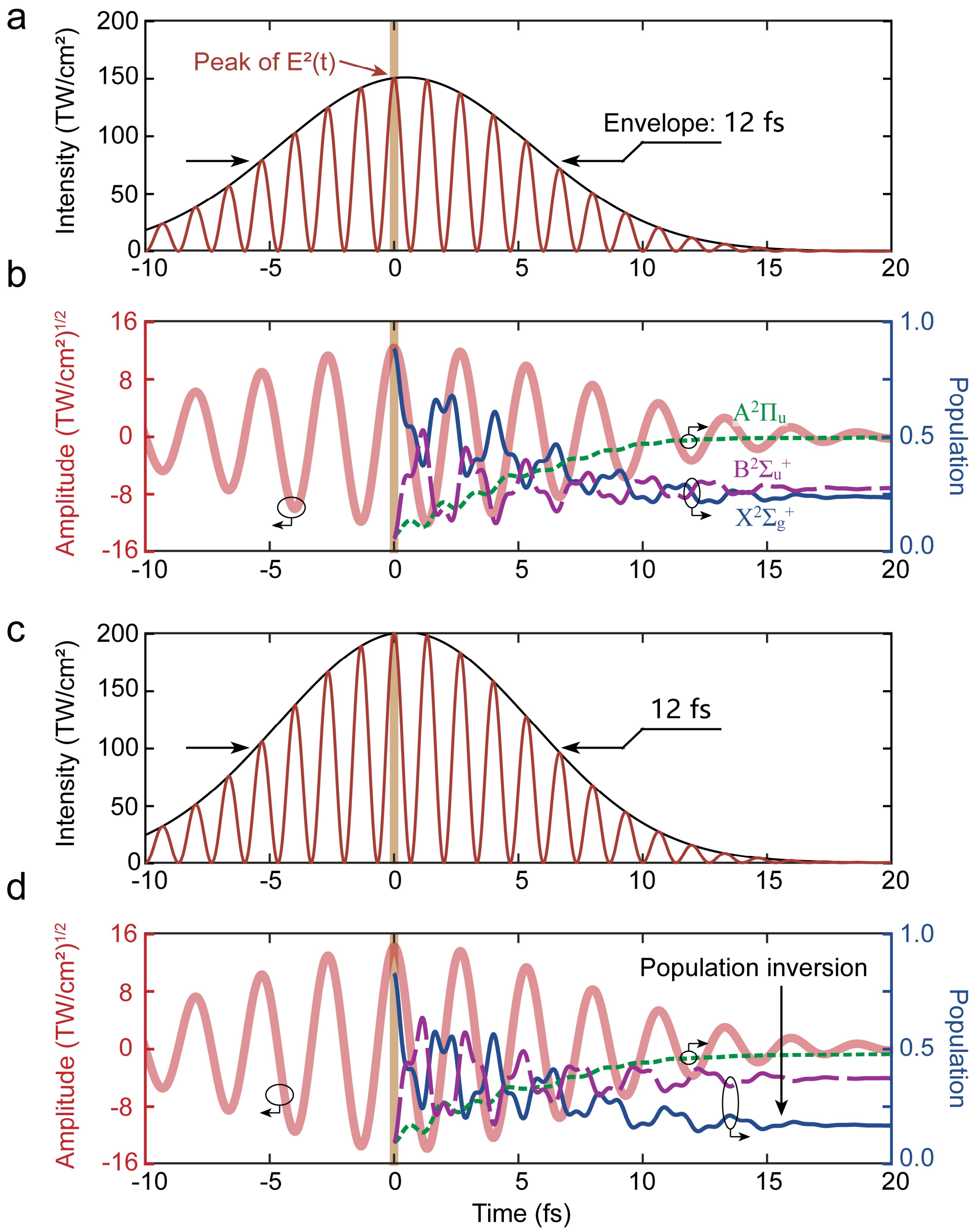

**Supplementary Fig. S5. Population dynamics of $N_2^+$ in 12-fs, 800-nm, Gaussian-shaped pulse. a,b** Pulse with peak intensity of 150 TW/cm$^2$ and corresponding population dynamics. **c,d** Pulse with peak intensity of 200 TW/cm$^2$ and corresponding population dynamics.

## Supplementary note 4: Dependence of $N_2^+$ lasing on pump energy

Supplementary Fig. S6 shows the the energy and conversion efficiency of the 391-nm and 358-nm $N_2^+$ lasing signals as a function of pump energy, extracted from the spectra in Fig. 3 of the main text. It should be noted that the spectra in Fig. 3 were recorded after the beam-splitting optics, and the corresponding $N_2^+$ lasing energies illustrated here are therefore slightly lower than those obtained direct at the HCF output port (Figs. 1d and 4 of the main text, and Supplementary Fig. S7).

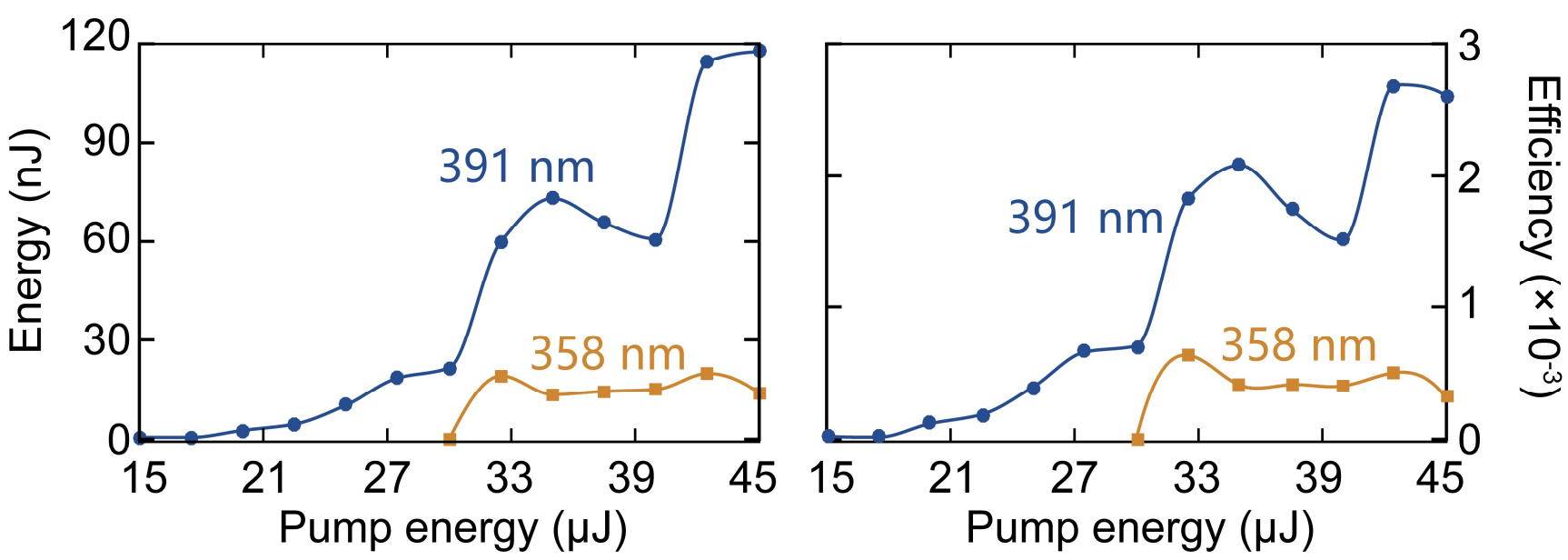

**Supplementary Fig. S6. Energy and conversion efficiency of $N_2^+$ lasing signals at different pump energies, corresponding to Fig. 3 in the main text.**

## Supplementary note 5: Wavelength-tunable dual-color ultrafast ultraviolet source

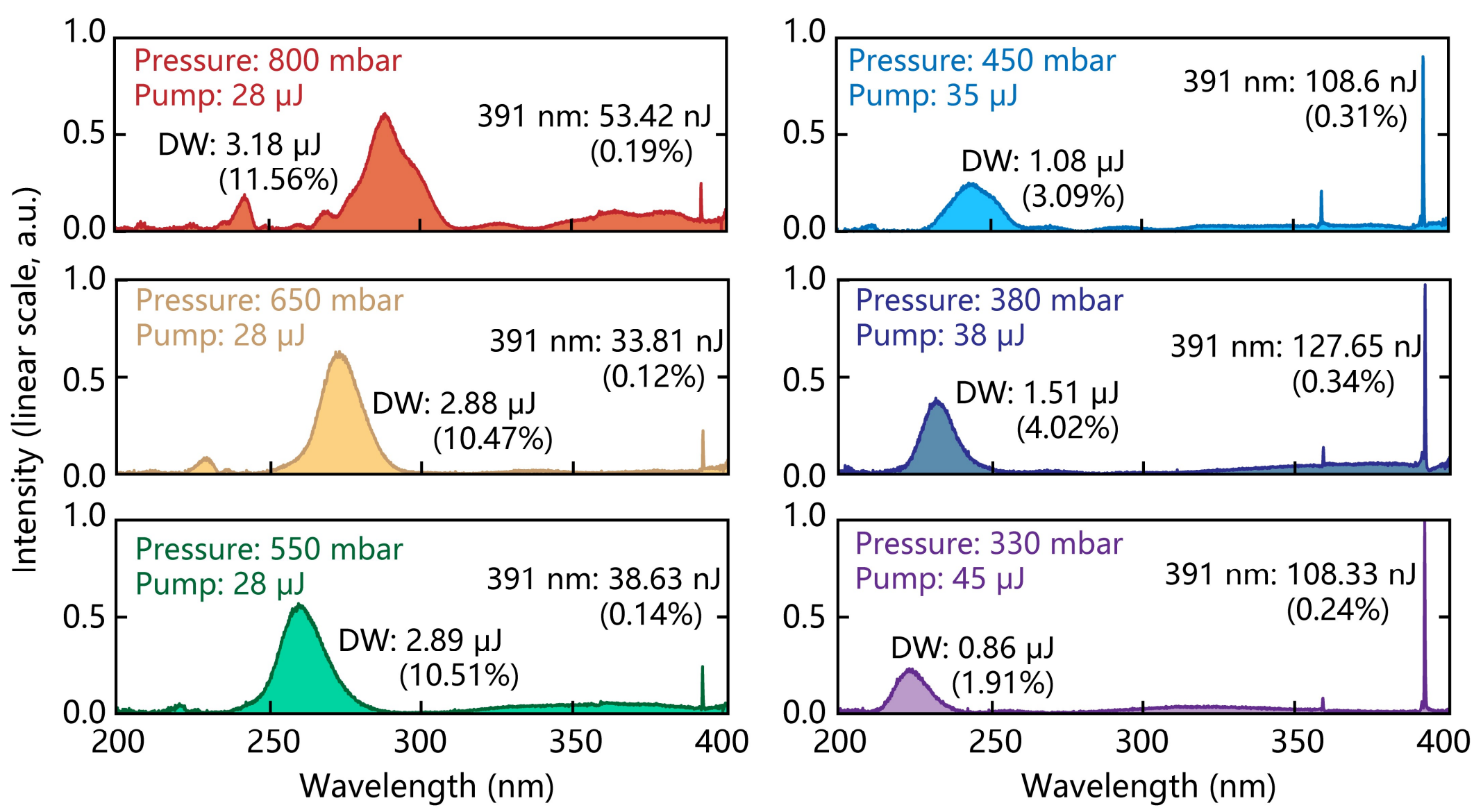

**Supplementary Fig. S7. Measured $N_2^+$ lasing and dispersive wave (DW) spectra from the tapered HCF, under different pressures and pump energies.**

## Supplementary references